\documentclass[pra,twocolumn,amssymb,showpacs]{revtex4}

\usepackage{appendix}
\usepackage{graphicx}
\usepackage{amssymb,amsmath} 

\def\B#1{\left(#1\right)}
\def\BB#1{\left[#1\right]}

\def\be{\begin{equation}}
\def\ee{\end{equation}}

\def\bee{\begin{equation*}}
\def\eee{\end{equation*}}

\def\for{\ \ {\rm for} \  }

\begin{document}

\title{Fidelity susceptibility of the quantum Ising model in the transverse field: The exact solution}

\author{Bogdan Damski} 
\affiliation{Los Alamos National Laboratory, Theoretical Division, MS B213,
Los Alamos, NM, 87545, USA \\
Institute of Physics, Jagiellonian University, Reymonta 4, 30-059 Krak\'ow, Poland} 
\begin{abstract}
We derive an exact closed-form expression for fidelity susceptibility of the quantum Ising
model in the transverse field. We also establish an exact one-to-one correspondence between fidelity 
susceptibility in the ferromagnetic and paramagnetic phases of this model. 
Elegant  summation formulas are  obtained as a by-product of these studies.
\end{abstract}
\maketitle

\section{Introduction}
Quantum phase transitions happen when a small variation in an external field
can fundamentally change ground state properties of a quantum  system
\cite{Sachdev,ContinentinoBook}. 
They provide some of the most striking examples of the  richness of many-body
quantum physics.
They can be studied in electronic \cite{Coldea}, cold atom \cite{cold},  
and cold ion \cite{ions} systems (see Refs.   \cite{Sachdev,LewensteinAdv}
for an overview of the field).
Traditional  condensed matter approaches to quantum phase transitions 
rely on the identification of the order parameter and studies of correlation functions \cite{Sachdev}. 
A different strategy has been recently proposed
by the quantum information community and is known as the fidelity approach 
to quantum phase transitions \cite{Zanardi,GuReview}. 

Fidelity is a popular concept in  quantum information science. 
It is  defined here as the overlap between two ground states
\bee
F(g,\delta)=\left|\langle\psi(g)|\psi(g+\delta)\rangle\right|,
\eee
where $|\psi(g)\rangle$ is a ground-state wave-function of some 
Hamiltonian $\hat H(g)$, $g$ is the external field whose variation 
induces a quantum phase transition, while $\delta$ is a small  shift of this
field.  Since the ground states fundamentally change
across the critical point, fidelity should have a minimum near the critical
point \cite{Zanardi}. This simple yet powerful observation is the basis 
of the fidelity approach to quantum phase transitions.

Recent studies suggest that fidelity is an efficient probe of quantum
criticality \cite{GuReview}. In particular, the minimum of fidelity near 
a critical point has been established in several models. The  scaling of 
fidelity with distance from the critical point $|g-g_c|$, the field shift $\delta$, and
the system size $N$ has been shown to encode the critical exponent $\nu$
characterizing the power-law divergence of the correlation length near the
critical point \cite{Zanardi,ABQ2010,Polkovnikov,MarekBodzioPRL}. 

Fidelity has been also shown to play a crucial role in  quantum 
phase transitions in quantum fields, where superpositions of ground states
from different phases are created \cite{BDPhysRep} (see also Refs. \cite{Ritsch,Morigi} for a different 
approach to quantum phase transitions in quantum fields). 
Moreover, fidelity has turned out to be
useful in studies of the dynamics  of quantum systems ranging 
from simple two-level  \cite{BDPRL2005} to  many-body ones 
(see e.g. Refs. \cite{Polkovnikov,Polkovnikov1} and Sec. V of Ref. \cite{MarekBodzioPRA}).
Its relevance  to the dynamics of decoherence of a central spin
coupled to an environment undergoing a quantum phase transition has been described in Ref.  \cite{BD_Dec}. 
There are two distinct regimes where fidelity can be studied. 

The first one 
corresponds to the limit where the field shift $\delta$ is kept constant and the system 
size $N\to\infty$. In this limit one observes Anderson
orthogonality catastrophe: Disappearance of the overlap of two ground states
in the thermodynamic limit \cite{Anderson1967}. Quite interestingly, while the Anderson's
seminal paper shows power-law decay of the overlap with the  system size in
a particular model that does not undergo a quantum phase
transition, there is an exponential with the system size decay of the overlap,
i.e., fidelity, near the generic quantum critical point 
\cite{MarekBodzioPRL,MarekBodzioPRA,SenPRB2012,Adamski2013} (see also 
Refs. \cite{Zhou,Zhou1}).

The second regime corresponds to the limit of the field shift $\delta\to0$  
taken at the constant system size $N$. In this limit fidelity is close to unity and it can
be approximated by the lowest-order nontrivial Taylor expansion 
\bee
F(g,\delta) \simeq 1 - \chi(g)\frac{\delta^2}{2},
\eee
where the linear in $\delta$ term vanishes due to normalization of the ground
states. The prefactor in the quadratic term, $\chi(g)$, is called fidelity susceptibility \cite{You2007}.
It has been recently intensively studied as a probe of quantum criticality
\cite{Zanardi,ABQ2010,Polkovnikov,GuReview}.

\begin{figure}[t]
\includegraphics[width=\columnwidth]{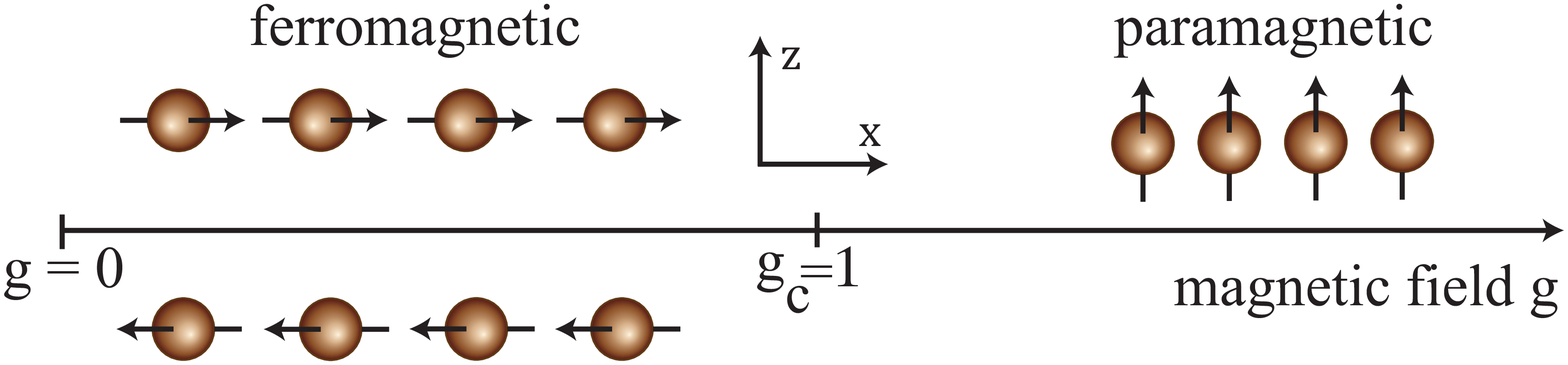}
\caption{(color online) Schematic of the phase diagram of the Ising model  in the transverse magnetic field.}
\label{fig1}
\end{figure}

\section{Model}
We will study fidelity susceptibility in the quantum Ising model in the
transverse field. This is a paradigmatic model of  quantum phase transitions:
All the basic concepts about both equilibrium \cite{Sachdev}  and non-equilibrium \cite{JacekAdv}
quantum phase transitions have been tested on this model. 
Its versatile experimental realization shall be possible in the nearest future in
cold ion setups \cite{ion_sim}.

The Hamiltonian that we study is given by 
\be
\hat H(g) = -\sum_{i=1}^N \left(\sigma^x_i\sigma^x_{i+1} + g\sigma^z_i\right),
\label{HIsing}
\ee
where $g$ is the  magnetic field and $N$ is the  number of spins.
Phase diagram of this  model  
reflects competition between the spin interactions and the  magnetic 
field (Fig. \ref{fig1}). The spin interactions try to align spins in the $\pm x$ direction, while the 
magnetic field  polarizes spins along the $z$ direction. For large enough magnetic
fields the system is in the paramagnetic phase, while for small magnetic fields it is in the
ferromagnetic phase. The two phases are
separated by the critical point $g_c=1$.

To find the ground states of the Hamiltonian (\ref{HIsing}), we assume that the number of spins is {\it even}
and proceed in the standard way  following notation from Ref. \cite{MarekBodzioPRA}. 
One obtains after some calculations 
\be
\begin{aligned}
&F(g,\delta)  = \prod_{k} \cos\frac{\theta_k(g+\delta)-\theta_k(g)}{2},\\ 
&\tan\theta_k(g) =\frac{\sin k}{g-\cos k},\\ 
&k=\pi/N, 3\pi/N,\dots,\pi-\pi/N.
\end{aligned}
\label{mainfid}
\ee
Fidelity susceptibility then reads 
\be
\chi(g)=\frac{1}{4}\sum_{k}\B{\theta'_k}^2
=
\frac{1}{4}\sum_{k} \frac{\sin^2k}{\B{g^2-2g\cos k+1}^2},
\label{chi}
\ee
where  the prime denotes the derivative with respect to $g$. Since 
fidelity susceptibility is symmetric around $g=0$, it is 
sufficient to consider magnetic fields $g\ge0$. Furthermore, it is useful to 
consider the vanishing magnetic field separately. 
We find with the help of  Ref. \cite{Ryzhik} that 
\bee
\chi(0) = \frac{N}{16}.
\eee
From now on, we assume that the magnetic field points in the $+z$ direction ($g>0$).
Moreover, we assume that $N\ge4$ excluding the trivial $N=2$ case.

There are at least four options for extracting information 
out of  Eq.  (\ref{chi}) -- note that this sum appears not only in the
studies of fidelity susceptibility, but also in the similar 
studies of quantum geometric tensors \cite{GeometricXY}, quantum Fischer information \cite{Invernizzi}, 
and  quantum adiabatic evolution \cite{Rezakhani}.
The first option is to perform the summation numerically, see e.g.
Ref. \cite{Zanardi}. This approach has obvious limitations.
The second option is to expand the summand in the Taylor series, see e.g. Refs. \cite{Zanardi,Polkovnikov,Invernizzi}.
This technique is necessarily approximate, it produces results whose range of
applicability is a priori unknown, and it cannot be deployed at arbitrary magnetic fields.
The third option is to factor out small systems,  replace 
$\sum_k$  by  $\frac{N}{2\pi}\int dk$, and calculate the integral, see e.g.
Refs. \cite{GuReview,Polkovnikov,Chen2008,GeometricXY,Rezakhani}. The main drawback 
of this approach is that it produces  results whose range of applicability can only be guessed. 
This range of applicability 
would be precisely known, had the ``remainders'' in the Euler-Maclaurin summation formula been 
studied \cite{Watson}, which is complicated  and  has {\it not} been yet done. 
In particular,  the replacement of the sum by the integral 
produces meaningless results near the critical point (see the  discussion below).   
The fourth, ultimate,  option is  to compute the sum  exactly analytically, which we will do below.

\section{Exact solution}
We start from an identity that can be found in Ref. \cite{Ryzhik},
\be
\sum_{k}\BB{\frac{\sin^2(k/2)}{\sinh(z)}+\frac{\tanh(z/2)}{2}}^{-1}=N\tanh\B{\frac{Nz}{2}},
\label{ryzhik}
\ee
valid for summations over the same $k$ as in Eq. (\ref{mainfid}).
Multiplying both sides of Eq. (\ref{ryzhik})  by $\tanh(z/2)$ and taking the
derivative of the resulting equation with respect to $z$, we obtain another  identity:
\begin{align}
&\sum_{k}\frac{\sin^2(k/2)}{\BB{\sinh^2(z/2)+\sin^2(k/2)}^2} =
f(z),\label{next}\\
&f(z)=\frac{N}{\sinh(z)}\frac{d}{dz}\BB{\tanh\B{\frac{Nz}{2}}\tanh\B{\frac{z}{2}}}.\nonumber
\end{align}
Next, we multiply both sides of Eq. (\ref{next}) by $\cosh^4(z/2)$ and
again differentiate the resulting equation with respect to $z$.
We obtain after some additional algebra 
\be
\begin{aligned}
\frac{d}{dz}\sum_{k} &\frac{\sin^2k}{\BB{\sinh^2(z/2)+\sin^2(k/2)}^2} = \\
&\frac{4}{\cosh^2\B{z/2}}\frac{d}{dz}\BB{\cosh^4\B{\frac{z}{2}}f(z)}.
\end{aligned}
\label{next2}
\ee
Then, we integrate Eq. (\ref{next2})  over $z$ from $0$ to $x$ getting
\be
\begin{aligned}
&\sum_{k} \frac{\sin^2k}{\BB{\sinh^2(x/2)+\sin^2(k/2)}^2} -2N^2+2N =\\
&\frac{2N\tanh(Nx/2)}{\tanh(x)}-N^2\frac{\cosh(Nx)}{\cosh^2(Nx/2)},
\end{aligned}
\label{fff}
\ee
where the left-hand-side is obtained with the help of 
\be
\sum_{k}\frac{1}{\sin^2(k/2)}=\lim_{z\to0}f(z)=\frac{N^2}{2},
\label{f3}
\ee
following from Eq. (\ref{next}). Eq. (\ref{fff}) can now be cast into the following 
simple form:
\be
\begin{aligned}
&\sum_{k} 
\frac{\sin^2k}{\BB{\sinh^2(x/2)+\sin^2(k/2)}^2} =\\
&\frac{N^2}{\cosh^2(Nx/2)}-2N\B{1-\frac{\tanh(Nx/2)}{\tanh(x)}}.
\end{aligned}
\label{moj}
\ee

Substituting $x = \ln g$ into Eq. (\ref{moj}) and comparing the resulting expression to Eq.
(\ref{chi}),  we find 
\be
\chi(g)= 
\frac{N^2}{16g^2}\frac{g^N}{\B{g^N+1}^2}+\frac{N}{16g^2}
\frac{g^N-g^2}{\B{g^N+1}\B{g^2-1}}.
\label{exactus}
\ee
This result  is exact and remarkably simple.  
In particular, it  works at and around the critical point, where the most interesting physics happens. 
For example, by rewriting 
\be
\frac{g^N-g^2}{g^2-1} = \frac{g^2}{g+1}\B{1+g+\dots+g^{N-3}},
\label{gie}
\ee
we see that Eq.  (\ref{exactus}) is regular at $g_c=1$.

We notice from  Eq.  (\ref{exactus}) that 
\be
g^2\chi(g)=\B{\frac{1}{g}}^2\chi\B{\frac{1}{g}},
\label{symmetry}
\ee
which  can be also  verified  from Eq.  (\ref{chi}). An analogical result was 
proposed for full fidelity in Ref. \cite{Zhou}. This symmetry reflects the
Kramers-Wannier duality of the Ising model \cite{Kramers}, which has not yet 
been discussed in the context of fidelity susceptibility.

Thus, $g^2\chi(g)$ is symmetric with respect to the
\be
\begin{aligned}
{\rm ferromagnetic} &\leftrightarrow {\rm paramagnetic}, \\
g &\leftrightarrow \frac{1}{g},
\label{trans}
\end{aligned}
\ee
mapping. Equation   (\ref{symmetry}) establishes  ferromagnetic/paramagnetic duality 
of fidelity susceptibility: All information about fidelity 
susceptibility is contained in one of the phases and can be uniquely mapped to the
other phase. We will now simplify Eq.   (\ref{exactus}).

We introduce another variable to properly organize the following discussion, 
\bee
y=N\ln g, \quad g=\exp\B{\frac{y}{N}}.
\eee
Its physical meaning is simple,
\bee
y\sim{\rm sign}(g-1)\frac{N}{\xi(g)},
\eee
where $\xi(g)$ is 
the correlation length of the infinite Ising chain in the transverse field $g$.
Note that we used the exact expression for the correlation
length, $\xi(g)\sim1/|\ln g|$, instead of the approximate one,  
$\xi(g)\sim 1/|g-1|$, valid near the critical point. 
The prefactors in both expressions are of the
order of unity and depend on the type of spin correlations used to define
the correlation length \cite{Pfeuty}.
The ferromagnetic/paramagnetic mapping (\ref{trans}) is equivalent to the
\bee
y\leftrightarrow-y
\eee
mapping. 

\begin{figure}[t]
\includegraphics[width=\columnwidth]{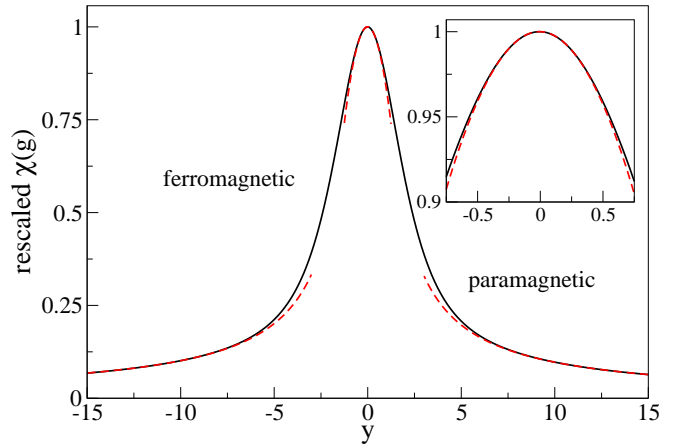}
\caption{(color online) Rescaled fidelity susceptibility. Solid black line: exact result (\ref{exactus}). 
Red dashed lines: approximations near and away from the critical point: Eqs.
(\ref{chi_near}), (\ref{chi_para}), and (\ref{chi_ferro}) without remainders. 
The inset shows enlargement of the central  part of the plot.
The system size $N=1000$. Fidelity susceptibility is rescaled by the 
$N(N-1)/32$ factor.  
}
\label{fig2}
\end{figure}

{\bf Away from the critical point}. We define that the system is away from the critical point 
when
\bee
|y|>1.
\eee
There are two advantages of the reparametrization of $\chi$ in terms of $y$. 
First, while $|g-1|$ is
bounded from above in the ferromagnetic phase, $|y|$ can be arbitrarily large
in both phases making it a good parameter to consider the far away limit.
Second,  it is also a convenient variable for
the actual calculations (see the Appendix).

We start by rewriting fidelity susceptibility on the paramagnetic side, $g>1$,
as
\begin{align}
\chi(g) &= \frac{N}{16g^2\B{g^2-1}} + R(g), \label{chi_para}\\
R(g)&=\frac{N}{16g^2\B{1+g^N}}\B{\frac{Ng^N}{1+g^N}-\frac{g^2+1}{g^2-1}},\label{chi_remainder} \\
\left|\frac{R}{\chi-R}\right|&\le 2e^2y\exp\B{-y+\frac{2(y-N)}{N}\Theta(y-N)}.\label{bound}
\end{align}
Such splitting of  fidelity susceptibility into the leading  term and 
remainder $R(g)$ is exact. While the leading term was   known before
\cite{GeometricXY,GuReview,Chen2008},
remainder (\ref{chi_remainder}) was unknown. 
Knowledge of the remainder is crucial for determining where $\chi(g)$  can be  approximated 
by  the  leading term. 
Bound (\ref{bound}), proven in the Appendix for $y\ge1$, 
shows that for $y\gg1$ the remainder is negligible  (see also Fig. \ref{fig2}).

The leading term was known before because it follows from the
above-mentioned replacement of the sum by the integral in Eq. (\ref{chi}). 
Note that without knowing the range of applicability of
this approximation, one may draw a wrong conclusion about singularity of fidelity susceptibility 
near the critical point, i.e., in the limit of $g\to g_c=1$ (see e.g. Ref. 
\cite{GeometricXY} where one of the diagonal elements of the quantum geometric
tensor is equal to fidelity susceptibility that we study).
Our result  shows why the singularity is absent:
The singularity of the leading term in Eq. (\ref{chi_para}) is exactly cancelled 
 by the divergent part of   remainder (\ref{chi_remainder}).
This can be verified using Eq. (\ref{gie}).
Thus, fidelity susceptibility near the critical point is 
regular rather than singular, which we will discuss below. 

Using  ferromagnetic/paramagnetic duality (\ref{symmetry}), we readily obtain 
fidelity susceptibility on the ferromagnetic side,  $0<g<1$,
\be
\chi(g) = \frac{\chi\B{1/g}}{g^4}=\frac{N}{16\B{1-g^2}} + \tilde R, \quad \tilde R=\frac{R\B{1/g}}{g^4},
\label{chi_ferro}
\ee
and  analogical remarks to the ones formulated to discuss Eqs. (\ref{chi_para})--(\ref{bound}) apply here. 
In particular, 
bound on $\left|\tilde R/(\chi-\tilde R)\right|$
is  the same as in Eq. (\ref{bound}) after replacing $y$ by $-y$.
Despite  differences
in the functional form of the leading terms on both sides of the
critical point -- Eqs. (\ref{chi_para}) and (\ref{chi_ferro}) --  the two expressions are in
fact two sides of the same coin due to symmetry (\ref{symmetry}).

{\bf Near the critical point}. We define that the system is near the critical point 
when  
\be
|y|<1.
\label{ynear}
\ee 

We start by noting  that fidelity susceptibility does not
have maximum at the critical point $g_c=1$. This is a consequence 
of  ferromagnetic/paramagnetic duality (\ref{symmetry})
 allowing us to write 
\bee
\chi(g)=\frac{h(g)}{g^2},\quad h(g)=h\B{\frac{1}{g}}.
\eee
From the  symmetry of $h(g)$, we  see that $h(g)$
has an extremum at $g_c=1$ (as it is  not the constant function here). 
This implies that $\chi(g)$ cannot have maximum  at $g_c$ due to the
$1/g^2$ factor shifting the maximum to  $g_m<g_c$.

To simplify the exact result for fidelity susceptibility near the critical
point, we Taylor-expand $h(g(y))$ in $y$ and keep the $1/g^2$ factor
unchanged. Such an approximation is welcome for two reasons. First, it is efficient 
because $h(g(y))$ is an even function of $y$ due to the $g\leftrightarrow 1/g$ symmetry of $h(g)$: 
odd expansion terms  vanish. Second, it 
 preserves   symmetry    (\ref{symmetry}).
We get 
\begin{align}
\chi(g) &= \frac{N(N-1)}{32g^2}\B{1-\frac{N+1}{N} \frac{(N\ln g)^2}{6} + r(g)}, \label{chi_near}\\
&0\le r(g) \le \frac{N}{N-1}\frac{\B{N\ln g}^4}{40},
\label{rbound}
\end{align}
where $N\ln g=y$ is a small parameter near the critical point (\ref{ynear}).
We see from bound (\ref{rbound})  that for $|y|\ll1$  remainder $r(g)$ is negligible;
$r(g)$ can be  obtained by comparing Eqs. (\ref{exactus}) and (\ref{chi_near}). 
The bound on $r(g)$ is derived in the Appendix. The lower
bound  is valid for any $y$, while the upper bound is valid for
$|y|\le\sqrt{10}$. 
The closest result to Eq. (\ref{chi_near}) was published in Ref. \cite{Invernizzi}. Apart from lacking the
remainder, it suffers from computational errors that we list in Ref. \cite{errors}.

Right at the critical point, Eq. (\ref{chi_near}) predicts
\be
\chi(g_c=1)=\frac{N^2}{32}-\frac{N}{32}.
\label{chic1}
\ee
This has to be compared to predictions of the scaling theory of fidelity
susceptibility proposing that the leading (in the system size) contribution 
to $\chi(g_c)$ is proportional to $N^2$ \cite{ABQ2010,Polkovnikov}.
Note that both the leading and the subleading term  is exactly 
captured by Eq.  (\ref{chic1}). We mention also that  the duality
symmetry implies that  
\bee
\chi'(g_c=1) = -2\chi(g_c=1) = -\frac{N^2}{16} + \frac{N}{16}\neq0,
\eee
which explicitly shows that there is no extremum of fidelity susceptibility 
at the critical point.

\begin{figure}[t]
\includegraphics[width=\columnwidth]{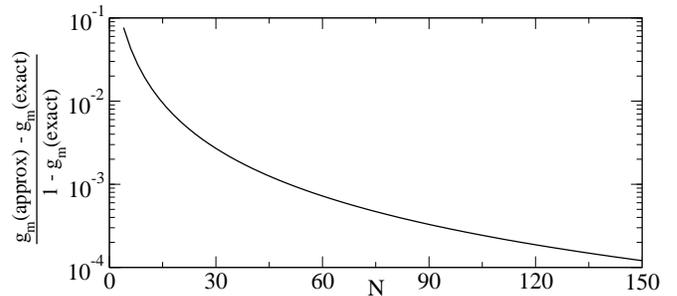}
\caption{Illustration of accuracy of the approximate expression for the
position of maximum of fidelity susceptibility. $g_m({\rm approx})$ is
given  by Eq. (\ref{gm_approx}), while $g_m({\rm exact})$ comes from numerical studies
of the exact expression (\ref{exactus}). For small
system sizes $N$ the shift of the maximum from the critical point
predicted by Eq. (\ref{gm_approx})  is 
off by a few percent and quickly decays with the system size. 
}
\label{fig3}
\end{figure}

Slightly away from the critical point, we enter the regime where the
term $\sim(N\ln g)^2$ starts to play a role. This term is responsible for shifting 
the  maximum of fidelity susceptibility away from the critical point. 
Indeed, forgetting about the remainder in Eq.  (\ref{chi_near})  
and solving equation $\chi'(g_m)=0$, we obtain (see also Fig. \ref{fig3})
\be
g_m = 1-\frac{6}{N^2}+\frac{6}{N^3} + O\B{N^{-4}}.
\label{gm_approx}
\ee
The limit of $N\gg1$ was assumed to simplify this result; analogical
calculation done on the exact expression (\ref{exactus}) predicts the same 
in the considered order.
This shows that indeed the maximum of fidelity susceptibility is located near the critical
point on the ferromagnetic side, which we proposed from symmetry (\ref{symmetry}).  
The maximum moves towards  the critical point  when $N$ increases.

Moving further from the critical point, we stay in the regime where the
remainder is still negligible, while the term $\sim(N\ln g)^2$ describes
decrease of fidelity susceptibility from the maximum (Fig. \ref{fig2}). 
Equation (\ref{chi_near}) shows then  how the scaling prediction 
near the critical point, $\chi\sim N^2$, gradually  breaks down. Finally, 
for $|N\ln g|=|y|\sim 1$ the remainder becomes non-negligible and the crossover 
to the far away limit  begins.

\section{Summary}

Summarizing, we have derived an exact closed-form expression for fidelity susceptibility
of the quantum Ising model in the transverse field -- the quantity that has been
intensively ``approximately''  studied in different contexts over the last couple of years 
\cite{GeometricXY,Invernizzi,Rezakhani,Zanardi,Polkovnikov,GuReview,Chen2008}.
Moreover, we have found an exact symmetry of fidelity susceptibility  
showing that all information about it  is contained in one of the phases and can be easily  
mapped to the other phase. This symmetry follows from the Kramers-Wannier
duality of the Ising model \cite{Kramers}.
These two results allow for an elegant
illustration  of the fidelity approach to quantum phase transitions 
in arguably the most important model of  a quantum phase transition.
They also pave the way for analytical
improvements of several earlier studies involving similar summations, see e.g. 
Refs. \cite{Invernizzi,Rezakhani}. Finally, they should
stimulate analytical investigations of other exactly solvable models.

\begin{center}
{\bf ACKNOWLEDGMENTS}
\end{center}

I would like to thank Adolfo del Campo for several useful suggestions  and Marek Rams
for deep insights about the duality symmetry. 
This work is supported by U.S. Department of Energy through the LANL/LDRD Program
and the Polish National Science Center grant DEC-2011/01/B/ST3/00512.

\appendix
\renewcommand{\thesection}{}
\section{}
We bound below  the remainders in Eqs. (\ref{chi_para}) and (\ref{chi_near}).

{\bf Away from the critical point}. 
We will bound here $\left|R/(\chi-R)\right|$. 
This is conveniently done after first replacing $g$ by $\exp(y/N)$
\bee
\frac{R}{\chi-R}  = 
N\frac{\exp(2y/N)-1}{\BB{\exp(y/2)+\exp(-y/2)}^2}
- \frac{1+\exp(2y/N)}{1+\exp(y)}.
\eee

Assuming that $y\ge1$ and $N>2$,
\bee
\left|\frac{R}{\chi-R}\right| \le N\frac{\exp(2y/N)-1}{\BB{\exp(y/2)+\exp(-y/2)}^2}
+ \frac{1+\exp(2y/N)}{1+\exp(y)},
\eee
after applying the standard inequality, $|a-b|\le |a|+|b|$. 
Note that it is sufficient to study only positive $y$, 
i.e., to focus on the paramagnetic side, thanks to the duality symmetry.

The subsequent ``bounding'' proceeds as follows 
\bee
\begin{aligned}
&N\frac{\exp(2y/N)-1}{\BB{\exp(y/2)+\exp(-y/2)}^2} 
     \\ \le&  N\B{\exp(2y/N)-1}\exp(-y) \\\le
      &\B{e^2-1}y\exp\B{-y+\frac{2\B{y-N}}{N}\Theta(y-N)}.
\end{aligned}
\eee
The first step above is obvious. To perform the second step, 
we use the following inequalities from Ref. \cite{Hardy1}
\bee
\begin{aligned}
a^r - b^r < r(a-b) b^{r-1} \for 0<r<1,\\
a^r - b^r < r(a-b) a^{r-1} \for r>1,\\
\end{aligned}
\eee
which are valid for positive and unequal $a$ and $b$ (equality happens 
for $r=0$, $r=1$, or $a=b$). We substituted $a=e^2$, $b=1$, and $r=y/N$ to bound $\exp(2y/N)-1$. 

Then, we bound 
\bee
\frac{1+\exp(2y/N)}{1+\exp(y)}\le \B{\exp(2y/N)+1}\exp(-y).
\eee
Next we combine the above results and proceed as follows. 

For $1\le y\le N$ we  get
\bee
\begin{aligned}
\left|\frac{R}{\chi-R}\right| 
   &\le y\exp(-y)\B{e^2-1+\frac{1+\exp(2y/N)}{y}}\\ &\le 2e^2y\exp(-y).
\end{aligned}
\eee

For $y\ge N$ we  get
\bee
\begin{aligned}
\left|\frac{R}{\chi-R}\right| 
   &\le y\exp(-y+2y/N) \B{\frac{e^2-1}{e^2}+\frac{1+\exp(-2y/N)}{y}}\\
   &\le 2y\exp(-y+2y/N).
   \end{aligned}
\eee

Combining  these two bounds we get 
\bee
\left|\frac{R}{\chi-R}\right|\le 2e^2y\exp\B{-y+ \frac{2\B{y-N}}{N}\Theta(y-N)}.
\eee
Equality in this bound is reached  only at $y=\infty$.

{\bf Near the critical point}. We will bound remainder $r$ using the following inequalities 
\bee
\begin{aligned}
&{\rm I:} \ \   \quad \tanh(x) >  \frac{x}{1+x^2/3}, \\
&{\rm II:} \  \ \quad \tanh(x) <  x,\\
&{\rm III:} \ \quad \tanh^2(x) >  x^2 -\frac{2x^4}{3},\\
&{\rm IV:}  \quad \tanh(x) <  x -\frac{x^3}{3} + \frac{2x^5}{15}, \\
&{\rm V:}  \ \quad \frac{1}{\tanh(x)} >  \frac{1}{x}+\frac{x}{3}-\frac{2x^3}{15}.
\end{aligned}
\eee
All of them are valid for $x>0$ of interest in our calculations
(equalities hold at $x=0$).

Inequality I follows from the discussion presented in Sec. 3.6.13 of Ref. \cite{Mitrinovic}.
Inequality II is proven by considering  
\bee
q(x)=x-\tanh(x) \quad \Rightarrow \quad q'(x)=\tanh^2(x).
\eee
Since $q(0)=0$ and $q'(x>0)>0$, we have $q(x>0)>0$, which establishes
inequality II. Inequality III follows from  
\bee
\tanh^2(x) - x^2 +\frac{2x^4}{3} > \frac{x^6\B{9+2x^2}}{3\B{3+x^2}^2}>0, 
\eee
where inequality I was employed to bound  $\tanh^2(x)$. Inequality IV is proven by considering
\bee
\begin{aligned}
&q(x) = x -\frac{x^3}{3} + \frac{2x^5}{15} -\tanh(x)\quad \Rightarrow \\
&q'(x) = \tanh^2(x) -x^2 +\frac{2x^4}{3}>0,
\end{aligned}
\eee
where the bound for $q'(x)$ follows from the proof of inequality  III. Since
$q(0)=0$ and $q'(x>0)>0$, we have $q(x>0)>0$, which establishes inequality
IV. Inequality V straightforwardly follows from inequality IV.

To bound remainder 
\bee
\begin{aligned}
r =&\frac{2N}{N-1}\frac{g^N}{\B{1+g^N}^2}+\frac{2}{N-1}\frac{g^N-g^2}{\B{1+g^N}\B{g^2-1}}+  \\
&\frac{N(N+1)}{6}\B{\ln g}^2-1,
\end{aligned}
\eee
 we replace $g$ by $\exp(y/N)$ and rewrite the resulting expression to the form 
\bee
\begin{aligned}
r=&-\frac{N\tanh^2(y/2)}{2(N-1)} + \frac{\tanh(y/2)}{(N-1)\tanh(y/N)}+\frac{N+1}{N}\frac{y^2}{6}-\\
&\frac{N}{2(N-1)}.
\end{aligned}
\eee
Note that the remainder is an even function of $y$  thanks to the duality symmetry. 

We bound the remainder  from above by bounding $\tanh^2(y/2)$, $\tanh(y/2)$, and
$\tanh(y/N)$ with inequalities III, IV, and I, respectively. It leads to  
\bee
r \le \frac{y^4}{720} \frac{18N^2-10+y^2}{N(N-1)}\le\frac{N}{N-1}\frac{y^4}{40}, 
\eee
where the last step is valid for $|y|\le\sqrt{10}$.

Next, we show that the remainder is bounded from below by zero by bounding 
$\tanh^2(y/2)$, $\tanh(y/2)$, and $1/\tanh(y/N)$ with inequalities II, I, and V,
respectively. This gives us 
\bee
\begin{aligned}
r \ge \frac{y^4}{120} \frac{5N^4-20N^2-96}{N^3(N-1)(12+y^2)} \ge 0
\end{aligned}
\eee
for 
$$
N\ge\sqrt{2+\frac{2}{5}\sqrt{145}}=2.6108\dots
$$


\begin{thebibliography}{37}
\bibitem[{err()}]{errors}
\bibinfo{note}{Sec. IIIB of Ref. \cite{Invernizzi} focuses on studies of
  quantum Fisher information at zero temperature, $G_J$. After setting $J=1$,
  $h=g$, and $L=N$ in Eq. (14) of Ref. \cite{Invernizzi}, we find that
  $\chi(g)=G_J(g)/4g^2$. $G_J$ is expanded in Ref. \cite{Invernizzi} in the
  Taylor series in $N(g-1)$ being similar to our $N\ln g$. To get the final
  result similar to our Eq. (\ref{chi_near}), one needs \begin{align*} &\sum_k
  \cot^2(k/2) = \frac{N^2}{2}-\frac{N}{2},\\
  &\sum_k\frac{\cot^2(k/2)}{\sin^2(k/2)} = \frac{N^4}{6}-\frac{N^2}{6},
  \end{align*} valid for summations over the same $k$ as in Eq.
  (\ref{mainfid}). For the first sum authors of Ref. \cite{Invernizzi} claim
  that there is an additional $O(N^0)$ term with respect to the above result,
  while the expression that they list for the second sum is incorrect. We got
  the first sum above from Eq. (\ref{f3}). The second sum is obtained after
  noting that it equals $\sum_k\sin^{-4}(k/2)-\sum_k\sin^{-2}(k/2)$, where
  $\sum_k\sin^{-2}(k/2)$ is given by Eq. (\ref{f3}), while $$
  \sum_k\sin^{-4}(k/2)=\frac{N^4}{6}+\frac{N^2}{3}. $$ This comes from
  multiplying Eq. (\ref{ryzhik}) by $1/\sinh(z)$, then calculating derivative
  of the resulting expression with respect to $z$, and finally, after
  rearrangement of terms, from taking the limit of $z\to0$.}
\end{thebibliography}

\end{document}